\documentclass[aps,prd,draft,showpacs]{revtex4}

\begin{document}

\title{RG Invariant Higgs Pole Mass from the $O(\hbar )$ Expansion}
\author{Chungku Kim}
\affiliation {Department of Physics, College of Natural Science, Keimyung
University, Daegu 705-701, KOREA}
\date{\today}

\begin{abstract}
We obtain the one-loop pole mass of the standard model Higgs which is
invariant under the beta functions of the MS\ scheme by following the
procedure in which the classical action in the broken symmetry phase was
obtaind by adding the $O(\hbar )$ expansion of the vacuum expection value
determined from the effective potential of the unbroken phase to the Higgs
field $H$.
\end{abstract}
\pacs{11.15.Bt, 12.38.Bx}
\maketitle



\section{Introduction}

The pole mass which plays an important role in the process where the
characteristic scale is close to the mass shell\cite{Narison-1}, was shown
to be infrared finite and gauge invariant\cite{Kronfeld}. The one-loop
relation between the scalar quartic coupling constant and other physical
quantities such as the Higgs pole mass, Fermi constant and the
electromagnetic coupling constant of the standard model(SM) was obtained in
the Feynman gauge\cite{Sirlin} and the expression of the pole mass of the SM
in terms of Lagrangian parameter up to two-loop order was obtained later\cite
{Jegerlehner} in general $R_{\xi }$ gauge. The physical quantities such as
the pole mass and the beta functions obtained in the minimal subtraction(MS)
scheme was used in determining the Higgs mass bound\cite{Kielanowski} or the
vacuum stability analysis\cite{stability}. Recently, the renormalization
group(RG) invariance of the pole mass was proved from the RG equation of the
two-point Green function\cite{Kim2}. However, the beta functions take
different forms when calculated in different renormalization schemes and in
many cases, these beta functions are obtained in the unbroken phase with the
MS scheme\cite{beta}. In this sense, it seems to be necessary to obtain the
RG invariant pole mass in the broken phase with the RG functions of the MS
scheme. Actually, in Appendix B of \cite{Jegerlehner} where the on-shell
renormalization was used, the $\mu $ dependent terms of the one-loop pole
mass $M$ of the SM have the a form 
\begin{equation}
M^{2}=m^{2}\{1+\hbar (\sum_{i,j} c_{ij}\frac{m_{i}^{2}}{m_{j}^{2}}\log (%
\frac{m^{2}}{\mu ^{2}})+\text{terms independent of }\mu )\}.
\end{equation}
In this case, in order to satisfy the RG invariance of the pole mass such as

\begin{equation}
\mu \frac{\partial M^{2}}{\partial \mu }+\beta _{\lambda }\frac{\partial
M^{2}}{\partial \lambda }+\beta _{m^{2}}\frac{\partial M^{2}}{\partial m^{2}}%
=0,
\end{equation}
we need the one-loop beta function containing the ratios of the masses as 
\begin{equation}
\beta _{m^{2}}=-2\hbar m^{2}\sum_{i,j}c_{ij}\frac{m_{i}^{2}}{m_{j}^{2}}.
\end{equation}
By noting that the $(mass)^{2}$ is given by the multiplication of the vacuum
expectation value(VEV) and the coupling constants, this implies that $\beta
_{m^{2}}$ contains the ratio of the coupling constants. It is evident that
the $\beta _{m^{2}}$ obtained in the MS scheme given in Eq.(16) below do not
contain this kind of term. Recently, we have shown that\cite{Kim1} if we
take the classical action of the broken symmetry phase by adding the vacuum
expectation value determined from the effective potential to the scalar
field in the unbroken phase, the resulting effective action of the broken
symmetry phase is invariant under the RG equation with the beta functions of
the unbroken phase and the resulting pole mass is RG invariant\cite{Kim2}.
In this paper, we will calculate the one-loop SM Higgs pole mass in terms of
the Lagrangian parameters in the Landau gauge which is invariant under the
beta functions in the MS\ scheme by following the procedure used in \cite
{Kim2} in case of the neutral scalar theory. In Sec.II, we will determine
the vacuum expectation value(VEV) of the SM by using the effective potential
obtained in the Landau gauge\cite{EP}. Then we calculate the two-point Green
function of the SM by using the classical action of the broken symmetry
phase obtained by substituting the field with the vanishing VEV. As a
result, we can check that the tadpole contribution cancels exactly with the
tadpole counterterm\cite{Taylor} and the resulting Higgs pole mass was
invariant under the beta function of the MS\ scheme. In Sec. III, we give
some discussion and conclusion.

\section{RG Invariant SM Higgs Pole Mass and the $O(\hbar )$ Expansion}

Consider the gauge-scalar part of the SM in the unbroken phase with the bare
classical action given by 
\begin{equation}
I_{B}=\int d^{D}x[\left| D_{\mu }\Phi _{B}\right| ^{2}+m_{B}^{2}\Phi
_{B}^{\dagger }\Phi _{B}+\frac{1}{6}\lambda _{B}(\Phi _{B}^{\dagger }\Phi
_{B})^{2}+L_{gf}+L_{gh}],
\end{equation}
where $L_{gf}$ and $L_{gh}$ are gauge fixing and the ghost terms and $%
D\equiv $ $4-2\varepsilon \ $is the dimension of the space-time and $\Phi $
is the scalar doublet 
\begin{equation}
\Phi =\left( 
\begin{array}{l}
\text{ \ }\Phi ^{+} \\ 
\frac{1}{\sqrt{2}}(H+\Phi _{0})
\end{array}
\right) ,
\end{equation}
with 
\begin{equation}
D_{\mu }\Phi _{B}=(\partial _{\mu }-\frac{i}{2}g_{B}\overrightarrow{\tau }%
\cdot \overrightarrow{W}_{B\mu }-\frac{i}{2}g_{B}^{\prime }B_{B\mu })\Phi
_{B}.
\end{equation}
The bare quantities can be expressed in terms of the renormalized quantities
such as 
\begin{equation}
m_{B}=m(1+\delta _{m^{2}}),\text{ }\lambda _{B}=\mu ^{2\varepsilon }\lambda
(1+\delta _{\lambda })\text{ and }\Phi _{B}=\Phi (1+\delta _{\phi }),
\end{equation}
and similarly for $g_{B},$ $g_{B}^{\prime },$ $W_{B}$ and $B_{B}.$ Then the
effective action that give the 1PI Feynman diagrams with the external Higgs
lines is given by 
\begin{equation}
\exp (-\frac{1}{\hbar }\Gamma (\phi ))=\int D\Phi \exp [-\frac{1}{\hbar }%
\{I_{B}(H+\phi )+J\Phi \}],
\end{equation}
and the corresponding effective potential $V(\phi )$ can be obtained by
putting $\phi $ as a constant. In case of the broken symmetry phase where
the renormalized mass $m^{2}<0$, the Higgs field H develops a VEV $v$ which
satisfies 
\begin{equation}
\left[ \frac{\delta V(\phi )}{\delta \phi }\right] _{\phi =v}=0.
\end{equation}
from which we can determine $v$ as power series in $\hbar $ ; 
\begin{equation}
v=v_{0}+\sum_{n=1}^{\infty }\hbar ^{n}v_{n}\text{ }(v_{0}^{2}=-\frac{6m^{2}}{%
\lambda }).
\end{equation}
Now, we choose the classical action with the Higgs field which do not have
VEV by substituting

\begin{equation}
\Phi =\left( 
\begin{array}{l}
\text{ \ }\Phi ^{+} \\ 
\frac{1}{\sqrt{2}}(H+v+\Phi _{0})
\end{array}
\right) ,
\end{equation}
into Eq.(4) and with the counterterms obtained in the unbroken phase of the
MS scheme. By this choice, $H$ has a vanishing VEV and hence the tadpole
diagrams cancel automatically\cite{Kim2,Taylor} as compared to the usual on
shell scheme\cite{on-shell} where the counterterms of parameters of the
classical action are determined from the no tadpole condition. The resulting
classical action contains not only the order $\hbar ^{n}(n>1)$ $\frac{1}{%
\varepsilon }$ divergent MS counterterms but also finite counterterms coming
from the $\sum_{n=1}^{\infty }\hbar ^{n}v_{n}$ given in Eq.(10). From
Eq.(8), the effective action in the broken symmetry phase $\Gamma _{BS}(\phi
)$ is related to $\Gamma (\phi )$ as 
\begin{equation}
\Gamma _{BS}(\phi )=\Gamma (\phi +v).
\end{equation}
It was shown that $\Gamma _{BS}(\phi )$ satisfies the RG equation with the
MS beta functions in the unbroken symmetry\cite{Kim1} and the resulting pole
mass is RG invariant\cite{Kim2}.

Now, let us apply the above procedure to SM and obtain the RG invariant
Higgs mass. For this purpose, we start from the renormalized one-loop
effective potential obtained in Landau gauge\cite{EP} in which the Goldstone
bosons are massless and the ghost terms couple only to the gauge bosons. The
one-loop effective potential is given by $V(\phi )=V_{0}(\phi )+\hbar
V_{1}(\phi )$ where 
\begin{equation}
V_{0}(\phi )=\frac{1}{2}m^{2}\phi ^{2}+\frac{1}{24}\lambda \phi ^{4},
\end{equation}
and 
\begin{equation}
(4\pi )^{2}V_{1}(\phi )=\frac{1}{4}H^{2}(\overline{\ln }H-\frac{3}{2})+\frac{%
3}{4}G^{2}(\overline{\ln }G-\frac{3}{2})+\frac{3}{2}W^{2}(\overline{\ln }W-%
\frac{5}{6})+\frac{3}{4}Z^{2}(\overline{\ln }Z-\frac{5}{6}),
\end{equation}
where 
\begin{equation}
H=m^{2}+\frac{\lambda }{2}\phi ^{2},\text{ }G=m^{2}+\frac{\lambda }{6}\phi
^{2}\text{, }W=\frac{g^{2}}{4}\phi ^{2}\text{, }Z=\frac{(g^{2}+g^{\prime 2})%
}{4}\phi ^{2}\text{ and }\overline{\ln }H=\ln \frac{X}{4\pi \mu ^{2}}+\gamma
,
\end{equation}
and satisfies the RG equation with the MS beta functions 
\begin{eqnarray}
\beta _{\lambda } &=&\mu \frac{d\lambda }{d\mu }=\frac{\hbar }{(4\pi )^{2}}%
(4\lambda ^{2}-9\lambda g^{2}-3\lambda g^{\prime 2}+\frac{9}{4}g^{\prime 4}+%
\frac{9}{2}g^{2}g^{\prime 2}+\frac{27}{4}g^{4})+\cdot \cdot \cdot , 
\nonumber \\
\beta _{m^{2}} &=&\frac{\mu }{m^{2}}\frac{dm^{2}}{d\mu }=\frac{\hbar }{(4\pi
)^{2}}(2\lambda -\frac{9}{2}g^{2}-\frac{3}{2}g^{\prime 2})+\cdot \cdot \cdot
,  \nonumber \\
\gamma &=&\frac{\mu }{\phi }\frac{d\phi }{d\mu }=\frac{\hbar }{(4\pi )^{2}}(-%
\frac{9}{4}g^{2}-\frac{3}{4}g^{\prime 2})+\cdot \cdot \cdot .
\end{eqnarray}
By using Eq.(9) and (13)-(15), we obtain 
\begin{equation}
v_{1}=\frac{1}{16\pi ^{2}\lambda v_{0}}\left[ \frac{3\lambda }{2}m_{H}^{2}\{%
\overline{\ln }(m_{H}^{2})-1\}+\frac{9g^{2}}{2}m_{W}^{2}\{\overline{\ln }%
(m_{W}^{2})-\frac{1}{3}\}+\frac{9(g^{2}+g^{\prime 2})}{4}m_{Z}^{2}\{%
\overline{\ln }(m_{Z}^{2})-\frac{1}{3}\}\right] .
\end{equation}
where $m_{H}^{2}\equiv m^{2}+\frac{\lambda }{2}v_{0}^{2}=-2m^{2},$ $%
m_{W}^{2}\equiv \frac{1}{4}g^{2}v_{0}^{2}=-\frac{3g2}{2\lambda }m^{2}$ and $%
m_{Z}^{2}\equiv \frac{1}{4}(g^{2}+g^{\prime 2})v_{0}^{2}=-\frac{%
3(g^{2}+g^{\prime 2})}{2\lambda }$ $m^{2}$. Now, the one-loop two-point
Green function $\Pi (p^{2})$ is given by 
\begin{equation}
\Pi (p^{2})=p^{2}+m^{2}+\frac{\lambda }{2}v_{0}^{2}+\hbar \Pi ^{(1)}(p^{2}),
\end{equation}
where $\Pi ^{(1)}(p^{2})$ is given by 
\begin{eqnarray}
\Pi ^{(1)}(p^{2}) &=&\begin{picture}(270,30) \put(15,-5){ \line(1,0){20}}
\put(28,3){\circle{16}} \put(25,15) {H} \put(45,0) {+ } \put(65,-5){
\line(1,0){20}} \put(78,3){\circle{16}} \put(70,15) {$\phi _{\pm,0}$}
\put(105,0) {+} \put(125,-5){ \line(1,0){20}} \put(138,3){\circle{16}}
\put(130,15) {$W _{\pm}$} \put(155,0) {+ } \put(175,-5){ \line(1,0){20}}
\put(188,3){\circle{16}} \put(186,12) {Z} \put(208,5){\line(1,0){4}}
\put(221,5){\line(1,0){8}} \put(237,5){\circle{16}}
\put(245,5){\line(1,0){8}} \put(233,15) {H} \put(233,-10) {H}
\put(260,5){\line(1,0){4}} \put(271,5){\line(1,0){8}}
\put(287,5){\circle{16}} \put(295,5){\line(1,0){8}} \put(280,18) {$\phi
_{+}$} \put(280,-10) {$\phi _{-}$} \end{picture}  \nonumber \\
&&\begin{picture}(270,30) \put(5,0){\line(1,0){4}} \put(16,0){\line(1,0){8}}
\put(32,0){\circle{16}} \put(40,0){\line(1,0){8}} \put(30,10) {$\phi _{0}$}
\put(30,-15) {$\phi _{0}$} \put(60,0){\line(1,0){4}}
\put(71,0){\line(1,0){8}} \put(87,0){\circle{16}} \put(95,0){\line(1,0){8}}
\put(80,10) {$W_{+}$} \put(80,-18) {$W_{-}$} \put(115,0){\line(1,0){4}}
\put(126,0){\line(1,0){8}} \put(142,0){\circle{16}}
\put(150,0){\line(1,0){8}} \put(138,10) {Z} \put(138,-18) {Z}
\put(170,0){\line(1,0){4}} \put(181,0){\line(1,0){8}}
\put(197,0){\circle{16}} \put(205,0){\line(1,0){8}} \put(190,10) {$W_{\pm}$}
\put(190,-15) {$\phi _{\mp}$} \put(220,0){\line(1,0){4}}
\put(231,0){\line(1,0){8}} \put(247,0){\circle{16}}
\put(255,0){\line(1,0){8}} \put(240,10) {Z} \put(240,-15) {$\phi _0$}
\put(267,0){+} \put(280,2){\line(1,0){20}} \put(285,-2) {\bf X} \end{picture}
\nonumber \\
&&\begin{picture}(270,45) \put(0,0) {+} \put(20,-8){\line(1,0){16}}
\put(28,-8){\line(0,1){5}} \put(28,5){\circle{16}} \put(24,15) {H}
\put(45,0) {+} \put(65,-8){\line(1,0){16}} \put(73,-8){\line(0,1){5}}
\put(73,5){\circle{16}} \put(66,18){$\phi _{\pm,0}$} \put(90,0) {+}
\put(115,-8){\line(1,0){16}} \put(123,-8){\line(0,1){5}}
\put(123,5){\circle{16}} \put(115,17){$W _{\pm}$} \put(135,0) {+}
\put(160,-8){\line(1,0){16}} \put(168,-8){\line(0,1){5}}
\put(168,5){\circle{16}} \put(165,15){Z} \put(180,0){+}
\put(195,-3){\line(1,0){16}} \put(203,-3){\line(0,1){10}}
\put(203,7){\circle*{4}} \put(220,0) {.} \end{picture}
\end{eqnarray}
By substituting Eqs.(10) and (11) into Eq.(4), we can obtain the
counterterms for the tadpole as 
\begin{equation}
\begin{picture}(20,20) \put(8,3){\circle*{4}}\end{picture} =m^{2}v+\frac{%
\lambda }{6}v^{3\text{ }}+m^{2}v(\delta _{m^{2}}+2\delta _{\phi })+\frac{%
\lambda }{6}v^{3}(4\delta _{\phi }+\delta _{\lambda })\simeq \hbar
\{-2m^{2}v_{1}+m^{2}v_{0}(\delta _{m^{2}}^{(1)}-2\delta _{\phi
}^{(1)}-\delta _{\lambda }^{(1)})\},
\end{equation}
and that for the two-point Green function 
\begin{equation}
\mathbf{X}=\frac{\lambda }{2}(v^{2\text{ }}-v_{0}^{2})+2p^{2}\delta _{\phi }%
\text{ }+m^{2}(\delta _{m^{2}}+2\delta _{\phi })+\frac{\lambda }{2}%
v^{2}(4\delta _{\phi }+\delta _{\lambda })\simeq \hbar \{\lambda
v_{0}v_{1}+2p^{2}\delta _{\phi }+m^{2}(\delta _{m^{2}}^{(1)}-10\delta _{\phi
}^{(1)}-3\delta _{\lambda }^{(1)})\}.
\end{equation}
By using the one-loop MS counterterms in the Landau gauge 
\begin{eqnarray}
\delta _{\phi }^{(1)} &=&\frac{\hbar }{16\pi ^{2}\varepsilon }\{\frac{9g^{2}%
}{8}+\frac{3g^{\prime 2}}{8}\},\text{ }\delta _{m^{2}}^{(1)}=\frac{\hbar }{%
16\pi ^{2}\varepsilon }\{\lambda -\frac{9g^{2}}{4}-\frac{3g^{\prime 2}}{4}%
\},\   \nonumber \\
\delta _{\lambda }^{(1)} &=&\frac{\hbar }{16\pi ^{2}\varepsilon }\{2\lambda -%
\frac{9g^{2}}{2}-\frac{3g^{\prime 2}}{2}+\frac{9g^{4}}{4\lambda }+\frac{%
9(g^{2}+g^{\prime 2})^{2}}{8\lambda }\},
\end{eqnarray}
we can see that the tadpole counterterm $%
\begin{picture}(20,10)
\put(8,3){\circle*{4}}\end{picture}$ cancels with the tadpole diagrams given
in last line of Eq.(19) completely as 
\begin{equation}
\frac{\lambda }{2}v_{0}A(m_{H})+(3-2\varepsilon )\{\frac{g^{2}}{2}%
v_{0}A(m_{W})+\frac{(g^{2}+g^{\prime 2})}{4}v_{0}A(m_{Z})\}+%
\begin{picture}(20,20) \put(8,3){\circle*{4}}\end{picture}=0,
\end{equation}
where 
\begin{equation}
A(m)=\int \frac{d^{D}q}{(2\pi )^{D}}\frac{1}{q^{2}+m^{2}}=-\frac{m^{2}}{%
16\pi ^{2}}(\frac{1}{\varepsilon }+1-\overline{\ln }m^{2}).
\end{equation}
Then the one-loop two-point Green function $\Pi ^{(1)}(p^{2})$ can be
written as 
\begin{eqnarray}
\Pi ^{(1)}(p^{2}) &=&\lambda v_{0}v_{1}+2p^{2}\delta _{\phi }+m^{2}(\delta
_{m^{2}}^{(1)}-10\delta _{\phi }^{(1)}-3\delta _{\lambda }^{(1)})  \nonumber
\\
&&+\frac{\lambda }{2}A(m_{H})+(3-2\varepsilon )\{\frac{g^{2}}{2}A(m_{W})+%
\frac{(g^{2}+g^{\prime 2})}{4}A(m_{Z})\}  \nonumber \\
&&-\frac{1}{2}\lambda ^{2}v_{0}^{2}B_{0}(p,m_{H},m_{H})-\frac{1}{6}\lambda
^{2}v_{0}^{2}B_{0}(p,0,0)-\frac{1}{4}g^{4}v_{0}^{2}F(p,m_{W})  \nonumber \\
&&-\frac{1}{8}(g^{2}+g^{\prime 2})^{2}v_{0}^{2}F(p,m_{Z})-\frac{1}{2}%
g^{2}G(p,m_{W})-\frac{1}{4}(g^{2}+g^{\prime 2})G(p,m_{Z}),
\end{eqnarray}
where $B_{0}(p,m_{1},m_{2})$ is the one-loop momentum integrals of Passarino
and Veltman\cite{Passarino} 
\begin{equation}
B_{0}(p,m_{1},m_{2})=\int \frac{d^{D}q}{(2\pi )^{D}}\frac{1}{%
(q^{2}+m_{1}^{2})((p+q)^{2}+m_{2}^{2})}=\frac{1}{16\pi ^{2}}\{\frac{1}{%
\varepsilon }+\int_{0}^{1}\overline{\ln }(\alpha m_{1}^{2}+(1-\alpha
)m_{2}^{2}+\alpha (1-\alpha )p^{2})d\alpha \},
\end{equation}
and $F(p,m)$ and $G(p,m)$ are defined by 
\begin{eqnarray}
F(p,m) &=&\frac{p^{4}}{4m^{4}}\{B_{0}(p,m,m)-2B_{0}(p,m,0)+B_{0}(p,0,0)\}+(%
\frac{p^{2}}{m^{2}}+\frac{1}{2})\{B_{0}(p,m,m)\}-B_{0}(p,m,0)\}-\frac{1}{%
2m^{2}}A(m)  \nonumber \\
&&+(\frac{5}{2}-2\varepsilon )B_{0}(p,m,m),
\end{eqnarray}
and 
\begin{equation}
G(p,m)=\frac{p^{4}}{m^{2}}\{B_{0}(p,m,0)-B_{0}(p,0,0)\}+p^{2}\{2B_{0}(p,m,0)%
\}-\frac{A(m)}{m^{2}}\}+m^{2}B_{0}(p,m,0)+A(m).
\end{equation}
Note that the $\frac{1}{\varepsilon }$ divergent terms of the two-point
Green function cancel with those from the counterterms for the two-point
Green function ( \textbf{X} ) and only the contribution of the finite term $%
\lambda v_{0}v_{1}$ of \textbf{X} remains.

Now, the Higgs pole mass $M^{2}$ is defined to be the pole of the two-point
Green function given in Eq.(18) so that 
\begin{equation}
\lbrack \Pi _{R}(p^{2})]_{p^{2}=-M^{2}}=0,
\end{equation}
It turns out that the functions $B_{0}(p,0,0)$ an$d$ $B_{0}(p,m,0)$ which
were present in the functions $F(p,m)$ and $G(p,m)$ vanishes in the pole
mass $M^{2}$ and we obtain 
\begin{eqnarray}
M^{2} &=&-2m^{2}+\hbar [\Pi _{R}^{(1)}(p^{2})]_{p^{2}=2m^{2}}  \nonumber \\
&=&m_{H}^{2}+\hbar m^{2}[(4\lambda -3g^{2}-g^{\prime 2}+\frac{9g^{4}}{%
2\lambda }+\frac{9(g^{2}+g^{\prime 2})^{2}}{4\lambda })\ln (\frac{m_{H}^{2}}{%
\mu ^{2}})+3\lambda b(m_{H})+(-2g^{2}+\frac{9g^{4}}{2\lambda }+\frac{%
2\lambda }{3})b(m_{W})  \nonumber \\
&&+\{-(g^{2}+g^{\prime 2})+\frac{9(g^{2}+g^{\prime 2})^{2}}{4\lambda }+\frac{%
\lambda }{3}\}b(m_{Z})-2\lambda \{\ln (\frac{m_{H}^{2}}{\mu ^{2}}%
)-1\}-(g^{2}+\frac{9g^{4}}{2\lambda })\{\ln (\frac{m_{W}^{2}}{\mu ^{2}})-1\}
\nonumber \\
&&-(\frac{(g^{2}+g^{\prime 2})}{2}+\frac{9(g^{2}+g^{\prime 2})^{2}}{4\lambda 
})\{\ln (\frac{m_{Z}^{2}}{\mu ^{2}})-1\}],
\end{eqnarray}
where 
\begin{equation}
b(m)=2\sqrt{4z-1}\tan ^{-1}(\frac{1}{\sqrt{4z-1}})+\ln (z)-2\text{ }(\text{ }%
z\succeq \frac{1}{4}\text{ })
\end{equation}
with $z\equiv \frac{m^{2}}{m_{H}^{2}}\succeq \frac{1}{4}$ for $%
m^{2}=m_{H}^{2},$ $m_{W}^{2}$ and $m_{Z}^{2}$. One can check that the pole
mass given in Eq.(30) satisfies the RG equation for the pole mass (Eq.(4))
with the MS beta function $\beta _{m^{2}}$ given in Eq.(16).

\section{Discussions and Conclusions}

In thisn paper, we have calculated the one-loop pole mass of the standard
model(SM) Higgs in the Landau gauge which is invariant under the beta
functions of the MS\ scheme. By adding the $O(\hbar )$ expansion of the VEV
obtained from the effective potential of the unbroken phase to Higgs field $H
$, we obtain the classical action in the broken symmetry phase which
contains both the $\frac{1}{\varepsilon }$ divergent counterterms coming
from the MS scheme and the finite counterterms from the $\hbar $ expansion
VEV. The tadpole diagrams vanishes automatically and the resulting pole mass
was invariant under the RG equation with the beta functions of the MS scheme
obtained in the unbroken phase. This procedure to obtain the RG invariant
pole mass in the broken symmetry phase can be extended to the case of the
gauge bosons. Then the scalar quartic coupling constant $\lambda $ can be
expressed by these pole masses which can provide another initial condition
for $\lambda $ in contrast to the previous result\cite{Sirlin} where $%
\lambda $ was obtained in terms of the Higgs mass, Fermi constant and the
electromagnetic coupling constant and this is under the investigation.

\end{document}